# The High Energy Light Isotope eXperiment program of direct cosmic-ray studies

**S. Coutu,**[a,1] **P.S. Allison,**[b] **M. Baiocchi,**[c] **J.J. Beatty,**[b] **L. Beaufore,**[d,b] **D.H. Calderón,**[b] **A.G. Castano,**[d] **Y. Chen,**[a] **N. Green,**[e] **D. Hanna,**[f] **H.B. Jeon,**[d] **S.B. Klein,**[g] **B. Kunkler,**[g] **M. Lang,**[g] **R. Mbarek,**[d] **K. McBride,**[b,d] **S.I. Mognet,**[a] **J. Musser,**[g] **S. Nutter,**[h] **S. O'Brien,**[f] **N. Park,**[c] **K.M. Powledge,**[d] **K. Sakai,**[d] **M. Tabata,**[i] **G. Tarlé,**[e] **J.M. Tuttle,**[d] **G. Visser,**[g] **S.P. Wakely**[d] **and M. Yu**[a]

[a] *Pennsylvania State University, Institute for Gravitation and the Cosmos, Dept. of Physics, 104 Davey Laboratory, University Park, PA 16802, USA*

[b] *The Ohio State University, Center for Cosmology and AstroParticle Physics, Dept. of Physics, Physics Research Building, 191 West Woodruff Avenue, Columbus, OH 43210, USA*

[c] *Queen's University, Dept. of Physics, Engineering Physics and Astronomy, Stirling Hall, 64 Bader Lane, Kingston, ON K7L 3N6, Canada*

[d] *University of Chicago, Enrico Fermi Institute, 933 East 56th Street, Chicago, IL 60637, USA*

[e] *University of Michigan, Dept. of Physics, Randall Laboratory, 450 Church Street, Ann Arbor, MI 48109, USA*

[f] *McGill University, Dept. of Physics, Ernest Rutherford Physics Building, 3600 University Street, Montreal, QC H3A 2T8, Canada*

[g] *Indiana University, Physics Dept., Swain Hall West, Bloomington, IN 47405, USA*

[h] *Northern Kentucky University, Dept. of Physics, Geology and Engineering Technology, Science Center 147, Highland Heights, KY 41076, USA*

[i] *Chiba University, Aerogel Factory Co., Ltd., 401 KCRC Building, 1-33 Yayoicho, Inage-ku, Chiba-shi, Chiba 263-0022, Japan*

*E-mail*: sxc56@psu.edu

ABSTRACT: HELIX is a new NASA-sponsored instrument aimed at measuring the spectra and composition of light cosmic-ray isotopes from hydrogen to neon nuclei, in particular the clock isotopes $^{10}$Be (radioactive, with 1.4 Myr lifetime) and $^{9}$Be (stable). The latter are unique markers of the production and Galactic propagation of secondary cosmic-ray nuclei, and are needed to resolve such important mysteries as the proportion of secondary positrons in the excess of antimatter observed by the AMS-02 experiment. By using a combination of a 1 T superconducting magnet spectrometer (with drift-chamber tracker) with a high-resolution time-of-flight detector system and ring-imaging Cherenkov detector, mass-resolved isotope measurements of light cosmic-ray nuclei will be possible up to 3 GeV/n in a first stratospheric balloon flight from Kiruna, Sweden to northern Canada, anticipated to take place in early summer 2024. An eventual longer Antarctic balloon flight of HELIX will yield measurements up to 10 GeV/n, sampling production from a larger volume of the Galaxy extending into the halo. We review the instrument design, testing, status and scientific prospects.

---

[1] Corresponding author.

# 1. Primary and Secondary Cosmic Rays

Galactic cosmic rays (CRs) are thought to be accelerated in diffusive shocks associated with supernova remnants (see, e.g., [1]). Their energy spectrum extends to at least a few times $10^{15}$ eV, following a power-law distribution $\sim E^{-2.6}$ at Earth, with $E$ the particles' kinetic energy. The spectral index of about −2.6 is thought to result from an $E^{-2}$ dependence at the source (in accordance to the venerable Fermi mechanism [2]), steepened by a factor of $E^{-0.6}$ owing to Galactic diffusion between the sources and the Earth. These *primary* cosmic rays are nuclei, typically p, He, C, N, O, Ne, Mg, Si, or Fe, having been produced in the aftermath of the Big Bang (p and He), or in *s*-process stellar nucleosynthesis for the heavier elements, and all readily available for injection into the Galactic accelerators.

Other nuclei are present, in smaller abundances, at the source and are accelerated as well, but some nuclei are conspicuously absent altogether from the list. In particular, Li, Be and B are disfavored by the energetics of stellar fusion processes. Instead, Li, Be and B can be produced as a result of spallation of more massive nuclei interacting with particles of the interstellar medium (ISM) during Galactic propagation. Such nuclei are *secondary* and serve as markers and tracers of Galactic propagation effects – the same effects responsible for the steepening of cosmic-ray spectra by $E^{-0.6}$ from sources to the Earth.

Cosmic antimatter production devolves similarly from secondary processes – antiprotons directly in the hadronic collisions, and positrons through the $\pi \to \mu \to e$ decay chain. The striking excess of cosmic positrons observed by the PAMELA [3] and AMS-02 [4] space-borne detectors, above baseline expectations from the CR secondary processes, remains to be explained and is the subject of intense speculation - such as postulating additional antimatter due to annihilating dark matter particles. Meanwhile, the current generation of space-borne instruments is producing a wealth of new CR measurements, revealing subtle features and departures from the simple power-law spectra described above [5]. The high-energy cosmic landscape is more complex than previously thought, and disentangling the role of the acceleration mechanisms from propagation effects can be challenging. It is therefore crucial to refine and extend all direct measurements of CRs, primaries and secondaries, up to the highest possible energies and with the largest possible statistics. This is a challenging task since the fluxes fall rapidly with energy.

Be and B secondary nuclei are readily produced in the spallation of a C or O (or more massive) nucleus interacting with a proton (H) or helium nucleus (He) in the ISM, but the reactions are complex [6]. For instance, beryllium nuclei can be produced as different isotopes in reactions such as $^{12}C + H \to {}^{9}Be$, $^{16}O + H \to {}^{9}Be$, $^{28}Si + H \to {}^{7}Be$, $^{12}C + He \to {}^{7}Be$, etc. but the inelastic cross sections for these reactions are not all known. For example, to account for 72% of all secondary Be nuclei produced, 17 different reaction channels must be taken into account. To account for 85% of the Be produced requires 63 reaction channels and to account for 91% of the secondary production requires 270 reaction channels. Nevertheless, the widely used CR diffuse emission production and transport codes such as GALPROP [7] incorporate all available information, and they can be used in turn to assess a multiplicity of CR phenomena. As semi-empirical models, they require the input of as many experimental results as possible. For example, the production of secondary nuclei and their transport to Earth are affected by the scale height $L$ of the Galactic halo out to which CRs propagate and interact, as well as a diffusion coefficient parameter $D_{0xx}$ that drives the propagation of the parent nuclei and their interaction daughters. For example, Fig. 1 (left), adapted from [8], illustrates the phase space of possible values of $L$ and $D_{0xx}$ allowable (between the dashed lines) [9] by using CR B flux measurements (specifically the proportion of B to C) and Be isotope measurements (specifically $^{10}Be/^9Be$), as available prior to precision measurements of these secondary nuclear fluxes. Using GALPROP calculations constrained by (here putative) high-statistics measurements of the energy evolution of the CR B/C and $^{10}Be/^9Be$ ratios, the degenerate band of the available phase space can be shrunk appreciably to the elliptical region.

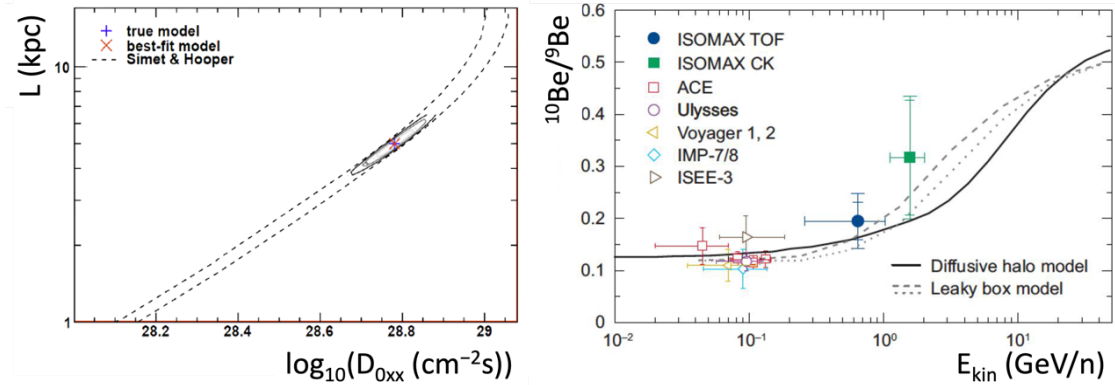

**Figure 1.** Left: parameter space of available values of CR Galactic halo scale height $L$ as a function of Galactic diffusion parameter $D_{0xx}$. The dashed lines represent a degenerate parameter region [9] allowable by the measured Be and B secondary nuclei fluxes prior to high precision measurements by the current generation of instruments. With new high-precision measurements of the energy evolution of B/C and $^{10}Be/^9Be$ ratios, the phase space becomes constrained to a region such as illustrated by the ellipses. Right: mass-resolved measurements of the $^{10}Be/^9Be$ ratio as a function of energy [10], as used in producing the parameter-space band in the left-hand plot. Three model predictions of the rise of the ratio with energy are indicated, but are not constrained by the available data.

Beryllium isotopes play a special role in assessing the physics of CR secondary production and propagation. Once generated, $^9Be$ nuclei are stable, while $^{10}Be$ will β decay with a half-life of 1.39 Myr. Given that the propagation times of CRs confined by Galactic magnetic fields, are typically on the order of 7-10 Myrs, the beryllium isotopes provide a suitable propagation clock. Additionally, the energy evolution of the $^{10}Be/^9Be$ ratio is expected to show an increase between 1 GeV/n and 10 GeV/n, owing to Lorentz time dilation becoming significant in this energy region. Thus while 1 GeV/n isotopes can reach the Earth from a region largely confined to the disk of the Galaxy, at higher energies the Galactic halo becomes progressively more highly sampled. Fig. 1 (right) shows mass-resolved measurements of the $^{10}Be/^9Be$ ratio as a function of energy [10], along with three model curves illustrative of the rise caused by the Lorentz time-dilated survival lifetime of $^{10}Be$ with increasing energy, allowing the secondary isotopes to reach Earth from a progressively larger volume of the Milky Way's halo. At present the available measurements do not constrain the models; this is the prime motivation for the HELIX program of light-isotope measurements.

## 2. High Energy Light Isotope eXperiment (HELIX)

Measuring and resolving isotopes of energetic light nuclei presents formidable experimental challenges. For a nucleus of charge $Ze$, mass $m$, momentum $p$, and velocity $\beta = v/c$, one can use a rigidity spectrometer to measure the rigidity $R$, given by

$$R = \frac{pc}{Ze} = \frac{\beta mc^2}{Ze\sqrt{1-\beta^2}}. \qquad (1.1)$$

The charge can be determined from the ionization energy loss by the nucleus traversing scintillator detectors, and the self-same scintillators can be deployed in layers some meters apart in a time-of-flight configuration to determine the velocity (practically up to about 1 GeV/n of incident CR kinetic energy). For higher energies (e.g., between 1 and 10 GeV/n), a ring-imaging Cherenkov (RICH) detector can be used to measure the velocity from the pattern of Cherenkov photon hits on a detector plane. They form a ring with a velocity-dependent radius since they are emitted in a cone of half-angle $\theta$ given by $cos\theta = 1/(\beta n)$, with $n$ the refractive index of the radiator material. Inverting Eq. (1.1) leads to the mass $m$ of the isotope, with a fractional mass resolution given by

$$\left(\frac{\Delta m}{m}\right)^2 = \left(\frac{\Delta R}{R}\right)^2 + \gamma^4 \left(\frac{\Delta \beta}{\beta}\right)^2 \qquad (1.2)$$

with $\gamma$ the Lorentz factor. For nuclei up to neon (Z = 10), to obtain, event-by-event, a mass resolution of better than 3%, up to ~3 GeV/n, HELIX is designed to measure $R$ with an uncertainty of ~2% and $\beta$ with an uncertainty of ~0.1%. For a description of the HELIX program, see [11].

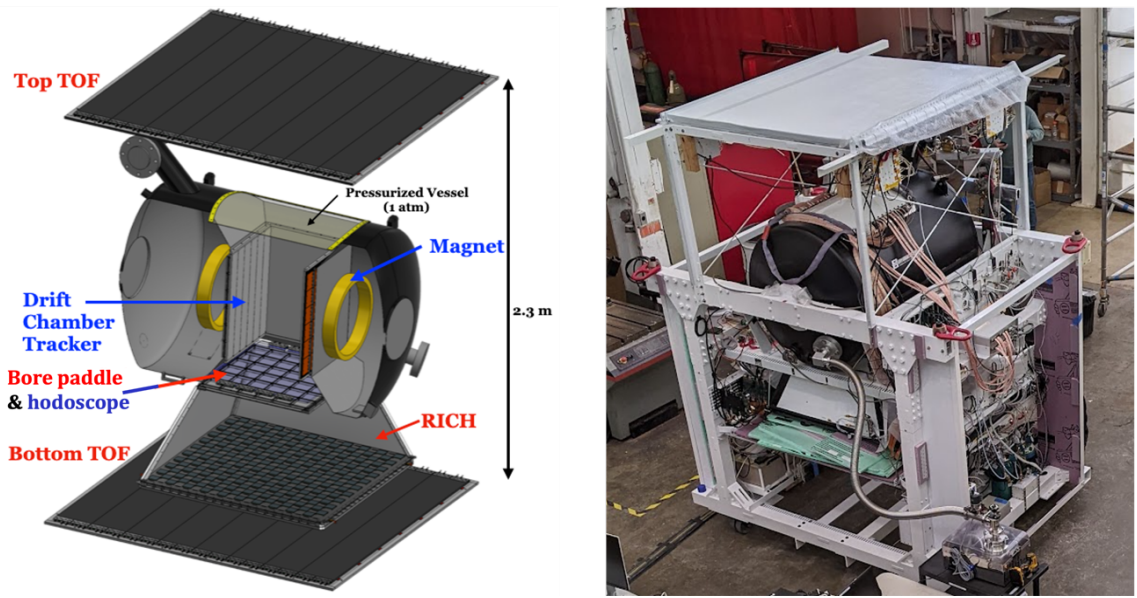

**Figure 2.** Left: a rendering of the HELIX detector arrangement. Right: the assembled HELIX instrument at a 2022 thermal-vacuum test at the NASA Neil A. Armstrong Test Facility.

The HELIX instrument is depicted schematically in Fig. 2 (left). A photo of the assembled instrument in Fig. 2 (right), taken during a thermal-vacuum test of the full instrument conducted at the NASA Neil A. Armstrong Test Facility in Sandusky, Ohio, in early 2022, is shown in Fig. 2 (right). The blue labels in Fig.2 identify the detector elements used for measuring $R$ in Eq. (1.1) and reconstructing the particle's track in three dimensions. The detector elements labeled in red are those used to trigger the instrument and to measure $Ze$ and $\beta$.

At the core of the instrument is a warm-bore, 1 T superconducting magnet. A drift-chamber tracker (DCT) is installed in the bore. It is enclosed in a pressure vessel to maintain the DCT at a pressure of 1 bar, independent of the ambient pressure, which will drop to a few mbar as the payload reaches the approximately 40 km float altitude. The magnet is enclosed in the black cryogenic vessel visible in Fig. 2 (right), with the curved top of the DCT pressure vessel in grey. A time-of-flight (TOF) system consists of plastic scintillators, 1 cm-thick, mounted at the top and bottom of the instrument and separated by 2.3 m. A bore-defining scintillator, also 1 cm-thick, is mounted just beneath the DCT to create an instrument trigger and provide a third sampling of the ionization energy deposited by the incident CR nucleus. The ionization energy deposit is used to measure $Ze$, and the time of transit between the scintillator layers is used to measure $\beta$ for nuclei with kinetic energies less than ~1 GeV/n. A RICH is installed between the bore paddle and the bottom TOF layer to measure $\beta$ for nuclei with kinetic energy between 0.3 and 10 GeV/n. The detector systems are described in the following sections.

## 2.1 Time-of-Flight System

The TOF is built from 1 cm-thick Eljen Technology EJ-200 scintillators [12]. The top and bottom TOF layers each consist of eight paddles, each $20 \times 160$ cm$^2$, and the bore paddle is a square with a 60.6 cm long edge. Figure 3 (left) shows a photograph of the bottom TOF layer. The scintillators are wrapped in three layers of extruded white PTFE (Polytetrafluoroethylene) film (0.003" thick) to improve collection of the scintillation light not captured by total internal reflection, and then further wrapped in three layers of black DuPont Tedlar (polyvinyl fluoride) film for light-tightness.

The top and bottom paddles are viewed at each end by 8 Hamamatsu S13360-6050VE silicon photomultipliers (SiPMs) [13], for a total of 16 per paddle. The bore paddle is viewed along two opposite sides by 16 SiPMs each (32 total), of the same model. These SiPMs have an active surface area of $6 \times 6$ mm$^2$ each, and are not affected by fringe magnetic fields from the magnet, so require no shielding. They feature a large dynamic range and high photo-detection efficiency, a fast (0.9 ns) risetime, and low transit-time skew. They operate with bias voltages of around 57V, far less than voltages used for conventional photomultiplier tubes. This effectively eliminates the problems of voltage breakdown in the low ambient atmospheric pressure at balloon altitude.

SiPMs are noisy, putting out dark noise rates on the order of 2 MHz at room temperature, but this can be mitigated by cooling. For the TOF SiPMs we use a passive scheme of thermally coupling the front-end board circuits (where the SiPMs are mounted) to the instrument frame via aluminum strips and bars, visible in Fig. 3 (left) as the rounded, white-painted strips at right. Further, each front-end board is equipped with a temperature sensor with which the operating bias voltage of the SiPMs is automatically compensated to maintain a constant gain average profile (averaged over 4 SiPMs ganged together). Despite the SiPM noise rate, the coincidence condition, requiring at least one hit from each of the top, bore and bottom scintillator layers results in a very clean event set, with muons at sea level being recorded at 25 Hz, with over 99% efficiency. The area and solid angle of this trigger define a geometric factor of 0.1 m$^2$sr. Although the SiPMs are rated for operation (or even storage) at no colder than −20 °C, during the thermal-vacuum testing the full TOF system was cooled down to about −40 °C without any adverse effects, so thermal effects are not anticipated to be of any concern.

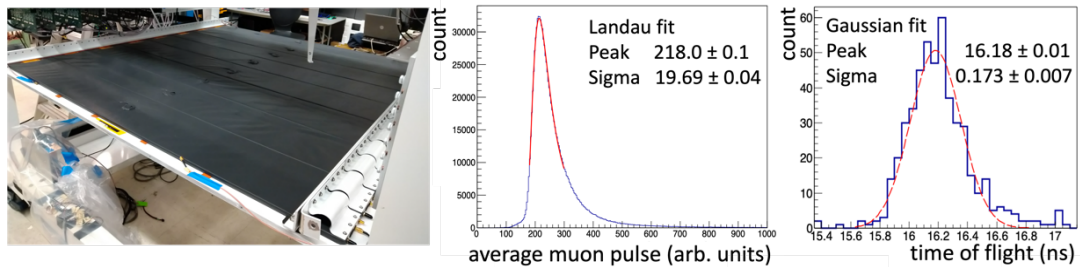

**Figure 3.** Left: bottom TOF layer, showing the 8 scintillator paddles. Middle: average integrated muon pulse signal (average of top, bore and bottom scintillators). The solid red line is a Landau fit, with a sigma/peak ratio of 9.0%. Right: distribution of time-of-flight values for near-vertical muons traversing the central regions of a top and bottom scintillator. The dashed red line is a Gaussian fit with a sigma of ~170 ps.

The SiPMs are mounted on carrier boards (4 per board), and are optically coupled to the scintillator faces using silicone optical cookies and Momentive two-component RTV615 optical silicone rubber. The SiPMs are passively summed in groups of four, and the signal readouts are sent to a custom readout board and separated into a slow charge signal and a separate fast timing signal. The timing signal is connected to a leading-edge discriminator, the output of which is connected to a time-to-amplitude converter sampled by a 14-bit analog-to-digital converter (ADC) running at 40 MSPS. This time-to-digital converter (TDC) configuration has a timing resolution better than 25 ps. Fig. 3 (right) shows the measured time-of-flight from the top to the bottom of the TOF system for ground-level muons traversing the central regions of a top and of a bottom scintillator paddle. The timing resolution achieved is ~170 ps; for Be isotopes, which have four units of charge and therefore generate more light, this is expected to improve by a factor of about 3.7 to the level of better than 50 ps. The slow output from the SiPM carrier board is shaped and then sampled with a 40 MSPS ADC channel. This signal is also used to provide the overall instrument trigger with outputs from two discriminator thresholds, allowing for independent low-charge and high-charge triggers, with optional pre-scaling on either. Fig. 3 (middle) is a distribution of average muon pulses from the top, bore and bottom scintillators. The resulting Landau shape has a width/peak ratio of 9.0%, and will allow the determination of the incident CR charge with a resolution of 0.1$e$. The ADC data on the readout board are received and packaged for transmission to the HELIX data acquisition (DAQ) by a Xilinx Artix-7 field-programmable gate array (FPGA), which also handles slow controls of the readout board including the SiPM bias system.

**2.2 Magnetic Spectrometer**

The magnet is a superconducting device built by Cryomagnetics [14] for the High Energy Antimatter Telescope (HEAT) balloon instrument [15]. It features two NbTi coils in a Helmholtz configuration, bracketing a warm bore of volume of 50 × 50 × 62 cm³, providing a 1 T central field at a current of 91.7 A. The aluminum cryostat holds 260 L of liquid helium (LHe), providing for a magnet hold time of about 6 days at float altitude. While previously operated within a large pressure vessel for high-altitude balloon flights, it was refurbished for HELIX to operate under stratospheric pressure conditions, and was successfully tested in a thermal-vacuum chamber simulating such an environment. The magnet's control, housekeeping and discharge systems were also replaced for HELIX with updated technology.

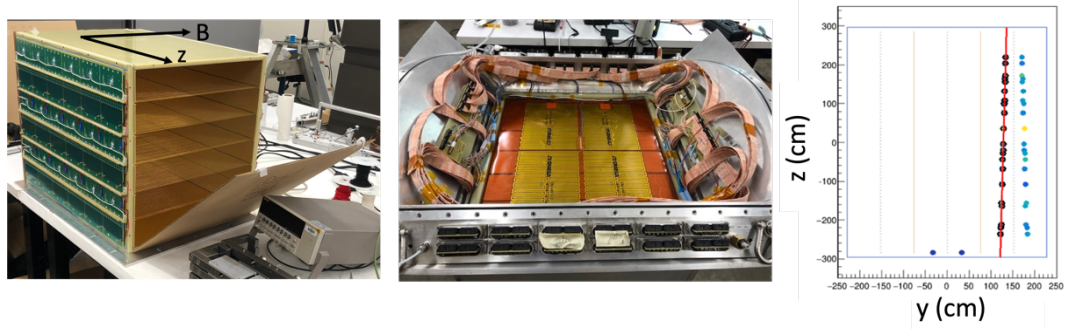

**Figure 4.** Left: DCT box on its side, prior to insertion into the magnet. The top lid is open, allowing the planes of sense wires, drift-field wires and field-shaping wires to be seen. The orientation of the magnetic field is indicated. Middle: view of the DCT inserted within its aluminum pressure vessel (but with the top off) installed inside the magnet bore. Right: muon track in a vertical yz plane with the best track fit line shown in red. The blue points are the left-right ambiguity options that have not been selected.

The DCT and its detailed specifications are described in [16]. It consists of a rectangular box filled with a $CO_2$-Ar (90:10) gas mixture, and crisscrossed with high-voltage cathode wires, field-shaping wires, and sense (tracker) wires. The ionization trail within the gas due to the CR transit is picked up by up to 72 sense wires (in the bending view of the spectrometer), each located no further than a 76 mm drift distance from the ionization cloud. The drift time to the sense wires is digitized and used to reconstruct a track, with a design goal of a 70 μm tracking resolution. The DCT is shown in Fig. 4 (left), resting on its side prior to insertion into the magnet bore and with its top lid open. The wire planes are visible. Figure 4 (middle) shows the DCT, within its pressure vessel, in place in the magnet bore, but with the pressure vessel lid removed.

The DCT consists of a box of internal dimensions $45 \times 45 \times 58$ cm$^3$ with walls made of 12.7 mm-thick FR-4 (glass-reinforced epoxy laminate material). Such thick walls provide structural support against deformation due to the wire tension, electrically insulate against breakdown of the high voltage on the wires to the aluminum pressure vessel. Strung across the volume of the chamber are 216 sense (tracking) wires made of moleculoy (non-magnetic nickel-chromium alloy, 4 Ω/mm). These are 20 μm-diameter, 48 cm-long, strung under a 0.19 N tension, and are arranged in three parallel vertical planes of 72 wires each. Figure 4 (right) displays DCT track hits by the passage of an undeflected (magnet off) CR-induced muon. The timing-derived hit positions result in a left-right ambiguity on either side of the sense wire plane. To resolve this, the wires are not perfectly aligned, but instead alternate wire positions are shifted by 0.3 mm, so that the hit positions can be analyzed using a Hough transform and the best solution can be identified (the black circles in the figure). A drift field is generated throughout the volume of the chamber using four cathode planes alternating with the three sense-wire planes. The two outer cathode planes (at the DCT walls) are formed by gold-plated circuit boards, while the inner two cathode planes each consist of 250 μm-diameter gold-plated aluminum wires. The cathode planes are held at a nominal –10 kV potential, which generates a strong electric drift field of 1.3 kV/cm. The need to enclose the DCT within a pressure vessel arises from the high voltage used, together with the need to circulate a drift gas throughout the chamber. To maintain a suitable set of operational conditions, the DCT temperature is made uniform and maintained with an array of heating pads attached to the outside of its walls. In addition to the cathode planes and sense wires, field-shaping wires (250 μm-diameter gold-plated aluminum) are used on a 4 mm pitch in the sense-wire planes to ensure a uniform drift field, and are held at a nominal –3 kV potential. This makes for a constant drift velocity of the ionization electron clouds. The use of $CO_2$ as a slow drift gas ensures that the drifting electrons do not accelerate to high velocities, minimizing Lorentz forces and thus improving the timing and tracking resolutions [15]. Each sense wire end is terminated at a chamber-mounted front-end (FE) board assembly. Sense wire signals are amplified and sent to ADC boards located outside of the pressure vessel via custom hermetic bulkhead feedthroughs. The ADC boards provide full waveform digitization of the 10 μs drift window following an event trigger, sampled at 80 MSPS at 12 bit resolution.

The sense wires have a nominal resistance of 1.8 kΩ, which is used to provide tracking positions in the non-bending view using charge division. To aid with the tracking resolution in this view, a scintillating-fiber hodoscope is inserted beneath the TOF bore scintillator, consisting of 600 square BCF-12 fibers (1 mm to a side), 1 m in length, procured from Saint-Gobain Crystals (now Luxium Solutions) [17]. The fibers are woven into a Delrin cookie with a position-multiplexed pattern mated to a Hamamatsu S14498 custom array of 64 individual SiPMs of size $6 \times 6$ mm$^2$, with 6336 micropixels on a 75 μm pitch (the same as used in the RICH, see below). This allows two or three fibers to be assigned to a single SiPM pixel, as described in detail in [18]. The positional accuracy of the transit point at the hodoscope is of order 1 mm.

## 2.3 Ring-Imaging Cherenkov Detector

At kinetic energies above 1 GeV/n, the timing resolution of the TOF is not sufficient to maintain the isotope mass resolution to better than 3%, so a RICH, described in detail in [18], is used. Figure 5 (left) shows the RICH prior to installation into the HELIX payload. At the top is an enclosure containing a $6 \times 6$ array of radiator tiles, each of dimensions $10 \times 10 \times 1$ cm$^3$. 32 of the tiles consist of custom-made hydrophobic aerogel material [19] with a nominal refractive index of 1.15 (Cherenkov threshold of 1 GeV/n, emission cone angle of 29º for fully relativistic particles), while four of the tiles consist of NaF with a refractive index of 1.33 (Cherenkov threshold of 0.45 GeV/n, emission cone angle of 41º for fully relativistic particles). The latter tiles provide an overlap for cross-calibration with the TOF measurement of $\beta$ up to 1 GeV/n. The aerogel tiles have been characterized via a comprehensive metrology and beam calibration program [20], yielding a measurement of their refractive index to a precision of 1 part in 10$^4$ on a 5 mm-pitch grid over their surface areas. A truncated pyramid, 50 cm-high and shown in Fig. 5 (left), provides the expansion volume for the Cherenkov light cone, which is incident upon a focal plane of 200 Hamamatsu S14498 SiPM arrays. These are deployed in a checkerboard configuration visible in Fig. 5 (middle). This covers only half of the detector plane, a cost-saving measure implemented for the initial HELIX flight.

The SiPM arrays are read out by front-end electronics boards through flexible PCB cables with a length of 70 cm, pictured in Fig. 5. Each front-end board contains 16 CITIROC 1A ASICs [21] and can process eight SiPM arrays. Each CITIROC amplifies its signal and routes it via two paths to fast and slow shapers which are used in generating a time stamp with a resolution better than 12.5 ns and multiplexed charge output, respectively. Figure 5 (right) shows a distribution of the gain-matched output of 12,697 individual SiPMs on the focal plane in response to low-intensity light flashes, showing the well-resolved photoelectron counts from the pedestal at left to 17 photoelectrons at right.

To lower the temperature of the SiPMs, and thus to suppress dark currents, a liquid cooling system is employed, consisting of thermoelectric devices, a pump, a radiator and a reservoir. Two Peltier thermoelectric modules (CP2-127-10-L1-W4.5, Laird Thermal Systems) are placed near two diagonal corners of the focal plane. The colder sides of the Peltier modules are thermally connected to ~65 cm-long cooling rails with heat pipes, installed along two opposite sides of the focal plane, as depicted in Fig. 5 (left).

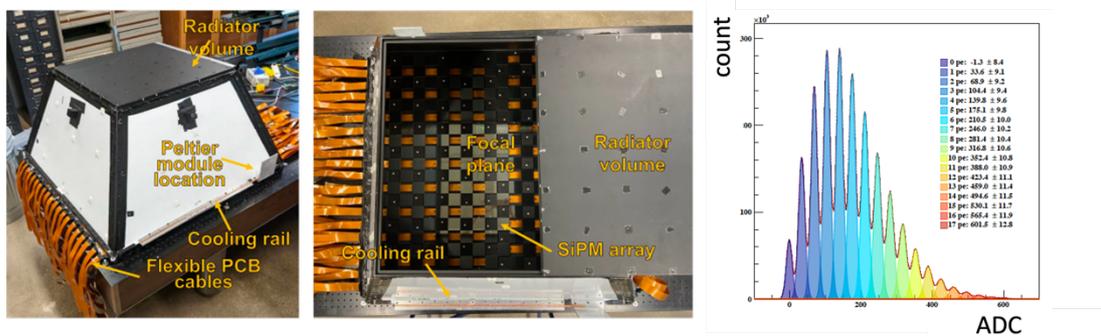

**Figure 5.** Left: RICH detector, showing the radiator volume enclosure at the top, the expansion volume (white truncated pyramid), and at the bottom the cables for readout of the SiPM photon sensors. A cooling rail and Peltier module location for SiPM temperature control are also seen. Middle: downward-looking view of the RICH with the radiator volume enclosure shifted, revealing the SiPM arrays at the focal plane. Right: distribution of the gain-matched output of 12,697 (of a possible 12,800) individual SiPMs on the focal plane in response to low-intensity light flashes, showing the well-resolved photoelectron counts from the pedestal at left to 17 photoelectrons at right.

### 3. Expected Performance

The first flight of HELIX by high-altitude balloon is anticipated to be launched from Kiruna, Sweden in early summer 2024, and to terminate over northern Canada approximately six days later. The payload will be recovered and refurbished for an eventual Antarctic balloon flight. The anticipated flight time along the Arctic route is a good match to the present magnet LHe hold time but an Antarctic flight will require some instrumental upgrades to lengthen the magnet hold time.

The HELIX instrument response has been investigated using detailed Monte Carlo simulations, and the anticipated beryllium mass isotope distribution is shown in Fig. 6 (left). With a mass resolution of 2.5% or so, the $^9$Be and $^{10}$Be isotope populations are resolved, and event-by-event particle identification is possible. Figure 6 (middle) shows the energy evolution of the $^{10}$Be/$^9$Be ratio as measured by the ACE/CRIS satellite and ISOMAX balloon instruments (repeated from Fig. 1 (right)), along with a prediction (purple curve) from a "leaky box" model [22], and anticipated HELIX measurements under that scenario. The HELIX points are shown in purple for a 14 day cumulative exposure (the first Arctic flight will have about half of the statistics assumed for those points), and for an eventual 28 day cumulative exposure, with Antarctic flights.

Also shown as red circles and black inverted triangles are reported analyses of AMS-02 data [23,24] (two different analyses of the same flight data). The trend of these AMS-02 points is at variance with the previous measurements or model predictions (shown in Fig. 1 (right)), and a model calculation is also shown (red curve) assuming a diffusive Galactic halo model with reacceleration [25]. It is worth pointing out that the AMS-02 points are obtained from template fits such as shown in Fig. 6 (right) for one kinetic energy bin, where $^9$Be and $^{10}$Be are not resolved event-by-event, but where instead the $^{10}$Be population causes a slightly broadened shoulder to the right of the $^9$Be population. There is clearly a need for the HELIX program to be brought to bear on the subject.

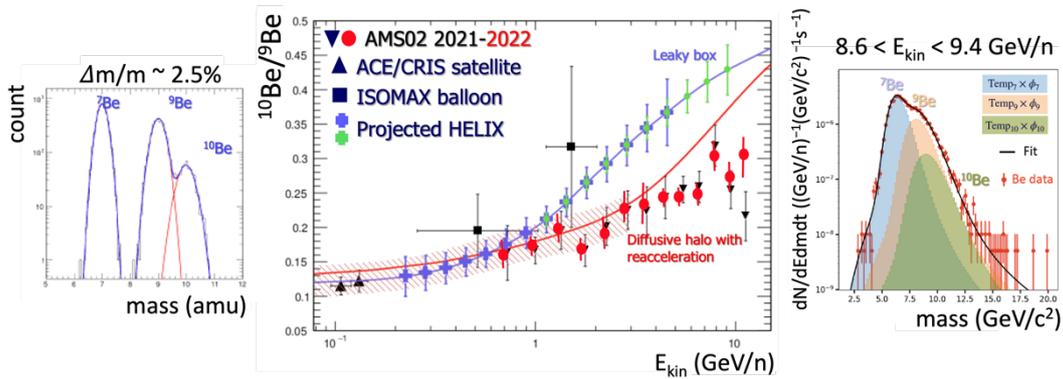

**Figure 6.** Left: anticipated beryllium isotope mass distribution for HELIX based on a detailed instrument Monte Carlo simulation. With a mass resolution of 2.5%, the $^9$Be and $^{10}$Be populations are resolved. Middle: anticipated energy evolution of the $^{10}$Be/$^9$Be ratio to be measured by HELIX (light purple points) with a 14 day exposure, assuming a "leaky box" model [22] (purple curve), and with a 28 day cumulative exposure (green points). Shown in red circles and black inverted triangles are two different analyses of AMS-02 data [23,24], along with a model [25] of a diffusive Galactic halo with reacceleration (red curve). Right: AMS-02 beryllium isotope mass distribution in one kinetic energy bin [23], where the $^{10}$Be and $^9$Be populations are not resolved event-by-event.

## Acknowledgments

This work is supported in the US by grant 80NSSC20K1840 from the National Aeronautics and Space Administration (NASA) and in Canada by grants from the Natural Sciences and Engineering Research Council (NSERC) and the Canadian Space Agency's Flights and Fieldwork for the Advancement of Science and Technology (FAST) program.